\documentstyle[12pt,amsfonts,amssymb,amscd]{report} \begin{document}
\begin{titlepage}
\pagestyle{empty}
\title{A new invariant for $\sigma $ models}
\author{Ioannis P.\ ZOIS\thanks{izois\,@\,maths.ox.ac.uk}\\
Mathematical Institute, 24-29 St. Giles', Oxford OX1 3LB}
\date{}
\maketitle
\begin{abstract}

 We introduce a \emph{new} invariant for $\sigma $ models (and foliations
more
 generally) using the \emph{even} pairing between K-homology and cyclic
homology. We try to calculate
 it for the simplest case of foliations, namely principal bundles. We end up by
 discussing some possible physical applications including quantum
gravity and M-Theory. In particular for
M-Theory we propose an explicit topological Lagrangian and then using
S-duality we conjecture on the existence of certain plane fields on $S^{11}$.\\

PACS classification: 11.10.-z; 11.15.-q; 11.30.-Ly\\
Keywords: Foliations; $\sigma $ models\\

\emph{To my brother Demetrios.}\\

\end{abstract}
\end{titlepage}

\section{Introduction and Motivation}

 In \cite {Z1} we proposed a Lagrangian density for the topological part of the
 non-sypersymmetric M-theory using Polyakov's flat bundle description
 of non-linear $\sigma $ models. The new key ingredient from the geometric
point of view was "characteristic classes for flat bundles". This idea
about using characteristic classes of flat bundles came from the
 definition of a new invariant for Haefliger structures \cite{Z2}.

 This invariant can be defined for \emph{any} foliation in general; as far as 
physics is concerned however (and this includes the case of M-theory
treated in \cite {Z1}), we are primarily concerned with a special kind of folia
tions, called \emph{flat foliations of bundles}. This is due to the fact that a
s Polyakov had noticed in \cite {polyakov},
 $\sigma $ models can be thought of as flat principal bundles (se
e \cite{Z1} for more details). Thus, hopefully, our invariant might be of some 
relevance whenever $\sigma $ models are met in physics.

 We organise this paper as follows: in section 2 we explain the strategy of the
 construction; in section 3 we provide all the details; in section 4 we
give the invariant formula; in section 5 we calculate this invariant for the si
mplest case of a principal bundle and in sections 6 and 7 
we discuss some possible applications in physics. In section 3 we
 review some of the techniques from non-commutative topology which we
 shall use for defining the invariant. All other sections contain new
 original material.

\section{Strategy}

\subsection{Instantons}

Let us recall some facts about instantons. We would like to think of our
invariant as an \emph{analogue of the instanton number for foliations}.

 We consider a principal bundle $(P,\pi,M,G)$, where $M$ is the base
manifold assumed to be compact and 4-dim for brevity, $G$ is SU(2) for simplici
ty, $P$ is the
total space of the bundle and $\pi$ is the projection. Assuming we have a
connection $A$ on $P$ with curvature $F$, then  the instanton number
ignoring constants is simply $\int _{M}F\wedge F$, i.e. the second Chern
number $c_2$ of the
bundle $P$.
 
 We would like to think of this number slightly differently: moreorless by
definition, any principal bundle $P$ over $M$ defines an element (K-class)
of the group $K^0(M)$ (we forget equivariant K-theory for simplicity).
Using the Chern-Weil homomorphism we get the Chern classes of $P$ which
belong to the \emph{cohomology} groups $H^{2*}(M)$. Considering the
(top dimensional) fundamental class $[M]$ of $M$ in the
\emph{homology} group $H_*(M)$  of $M$ and taking the \emph{pairing}
between homology and cohomology, which in this case is just integration
over $M$, we get the instanton number. We can consider the Chern-Weil
homomorphism from $K^0(M)\rightarrow H^{2*}(M)$ as a "black box" and
forget
all about cohomology for the moment; then the instanton number will be the
result from pairings between K-theory and (singular) homology $H_*(M)$.

 Our construction since we are dealing with foliations (more
accurately with the space of leaves of foliations) which provide a
good example of non-commutative topological spaces, immitates the above
picture: to each foliation we can associate a homology class which will be
the analogue of the fundamental class $[M]$ above; this class however will
belong to an appropriate homology theory called \emph{cyclic homology} and
it is called \emph{transverse fundamental class} of the foliation.
Moreover one can also construct a class in \emph{K-homology}, being the
analogue
of K-theory for our purpose. Then we use a formula for pairings between
cyclic homology and K-homology to get our result.

\subsection{Non-commutative Topology}

 In this subsection we would like to mention briefly what non-commutative
topology is about. As its name suggests, this is one aspect of
non-commutative geometry. Non-commutative geometry has appeared in physics
literature some years ago mainly through the so called "quantised
calculus". Anyway, the starting point of non-commutative topology is the
fact that given any compact Hausdorff space $X$ say, the commutative
($C^*$)-algebra $C(X)$ of complex valued functions defined on $X$ can
capture all the toplogical information of the space $X$ itself; in fact
$X$ and $C(X)$ are completely equivalent, one can be uniquely constructed
by the other. Conversely, given any commutative algebra $A$, say, there
exists a compact Hausdorff space $X$ say, (called the spectrum of $A$)
"realising" the commutative algebra $A$. Realising means that the
commutative algebra $C(X)$ of complex valued functions on $X$ is
essentially the algebra $A$. In mathematics terminology one says that the
categories of compact Hausdorff spaces and commutative $C^*$-algebras are
equivalent. This is the so called \emph{Gelfand's theorem}.

 We know however that there exist non-commutative $C^*$-algebras as well.
The natural question then is whether one can find a "topological"
realisation for them just like for the commutative ones. We are looking
for a non-commutative analogue of Gelfand's theorem. This question is not
fully answered in mathematics, it is related to the famous Boum-Connes
conjecture. There are some things already known in mathematics and these
are related to foliations. This is what we shall be using extensively in
this paper. The appropriate framework is that of K-theory and various
homology theories.

 During the '70s mathematicians (Baum, Douglas, Kasparov and others)
developed a K-theory for arbitrary $C^*$-algebras (commutative or not)
 and it is a well-known theorem due to Serre and Swan that in the
commutative
case this K-theory reduces to Atiyah's original topological K-theory.
Moreover in the 80's mathematicians (Connes, Loday, Quillen and others)
developed a homology theory called \emph{cyclic homology} for arbitrary
algebras which again in the commutative case gives in the limit the usual
simplicial homology. So non-commutative topology, in terms of K-theory and
various homology theories gives a generalisation of ordinary topology
through Gelfand's theorem. A good example of a non-commutative topological
space is the space of leaves of a foliation (see below for definitions).
In general quotients of ordinary topological spaces by discrete groups
give non-commutative (abreviated to \emph{"nc"} in the sequel) spaces.
 Good textbooks are \cite{wegge} and  \cite{loday} for an
introduction on K-theory of $C^*$-algebras and cyclic homology
respectively.

\subsection{The Invariant}

 In order to construct our invariant for \emph{any} foliation \cite{Z2},
we use
 some
ideas from non-commutative geometry \cite{connes1},
\cite{connes2}, \cite{connes3}, \cite{conness}. The strategy is as
follows: given any foliation
 $F$ 
of a manifold $V$, namely an integrable subbundle $F$ of $TV$, one can
associate to it another manifold $\Gamma (F)$, called the \emph{graph} (or
\emph{holonomy groupoid}) of the foliation introduced in \cite{win}. This is of
 dimension
$dimV+dimF$. Using the complex line bundle $\Omega ^{1/2}(\Gamma (F))$
of $1/2$-densities defined
on $\Gamma (F)$, we consider the set (actually vector space) of smooth
sections of this line
bundle equipped with a $*$ product, thus obtaining an algebra. We then
complete this algebra in a "minimal" manner (in standard $C^*$-algebra
theory this is called the \emph{reduced} $C^*$-algebra completion),
thus we obtain a $C^*$-algebra denoted $C^{*}(F)$ which is naturally
associated to our original
foliation $F$. From now on one can forget the original foliation $F$ of
$V$
alltogether and concentrate on its corresponding $C^*$-algebra $C^{*}(F)$.
We are
interested in the $K_0$ group of $C^{*}(F)$ and in its cyclic homology
groups.  If
we pick a metric $g$ on the transverse bundle $t$ of
$F$ we can  construct in a natural way a $C^{*}(F)$-module $E(F)$,
thus obtaining a class $[E(F)]$ in $K_0(C^{*}(F))$. Moreover to our
foliation one can associate in a natural way a cyclic cocycle $[F]$ in the
q-th cyclic homology group of the $C^*$-algebra $C^*(F)$, called the
\emph{fundamental transverse cyclic cocycle} of the foliation, where $q$
is the codimension of the foliation $F$. Then we use the even
pairing between K-homology and cyclic homology in this case, namely we
consider the pairing
$$\langle [E(F)],[F]\rangle :=(m!)^{-1}(F \# Tr)(E(F),...,E(F))$$
as was firstly introduced in the abstract algebraic context in \cite{connes3}. Hence we
obtain a \emph{complex number} as a result from the above pairing and
this complex number characterises our original foliation $F$.

\section{The constructions in detail:}

\subsection{Foliations}

 Let $V$ be a smooth manifold and $TV$ its tangent bundle. A smooth
subbundle $F$ of $TV$ is called \emph{integrable} iff one of the following
equivalent conditions is satisfied:

1. Every $x\in V $ is contained in a submanifold $W$ of $V$ such that

$$T_y(W)=F_y$$
where $T_y$ denotes the tangent space over $y$.

2. Every $x\in V$ is in the domain $U\subset V$ of a submersion
$p:U\rightarrow {\bf R}^q$ (q=codim $F$) with $F_y=ker(p_*)_y\forall y\in U$.

3. $C^{\infty }(V,F)={X\in C^{\infty }(V,TV); X_x\in F_x \forall x\in
V}$ is a Lie subalgebra of the Lie algebra of vector fields on $V$.

4. The ideal $J(F)$ of smooth differential forms which vanish on $F$ is
stable under differentiation:$d(J)\subset J$\\

 The condition 3. is simply Frobenius' Theorem and 4. its dual.\\

\emph{Example:}

 Any 1-dimensional subbundle $F$ of $TV$ is integrable, but for $dimF\geq
2$ the condition is non-trivial; for instance if $V$ is the total space
of a principal bundle with compact structure group, then we know that
the subbundle of \emph{vertical} vectors is always integrable, but the
\emph{horizontal} subbundle is integrable iff the connection is \textbf{flat}.\\
We shall make extensive use of this fact in this piece of work.

 A \emph{foliation} of $V$ is given by an integrable subbundle $F$ of $V$. The
\emph{leaves} of the foliation are the maximal connected submanifolds
$L$ of $V$ with $T_x(L)=F_x \forall x\in L$, and the partition of $V$ into
leaves $V=\cup L_a$ where $a\in A$ is characterised geometrically by
its "local triviality": every point $x\in V$ has a neighborhood $U$ and
a system of local coordinates $(x^j)$, j=1,...,dimV which is called
\emph{foliation chart} , so that the partition of $U$ into connected
components of leaves, called \emph{plaques} (they are the leaves of
the restriction of the foliation on $U$), corresponds to the partition
of ${\bf R}^{dimV}={\bf R}^{dimF}\times {\bf R}^{codimF}$ into the parallel affine
subspaces ${\bf R}^{dimF}\times pt$.

Very simple examples indicate that the leaves $L$ may \textbf{not} be compact even
if the manifold $V$ is and that the space of leaves $X:=V/F$ may \textbf{not} be
Hausdorff for the quotient topology. The "rational torus" is such an example.\\

Throughout this paper we would mainly restrict our attention to \emph{two
special kinds of
foliations:}
  we  consider a principal bundle $P$ with structure (Lie) group $G$
(assumed compact and connected) over a compact manifold $M$. The total space $P
$ has automatically a foliation
induced by the fibration: the leaves are the fibers which are
isomorphic to the structure group $G$ and the space of leaves is just
the base space $M$ with its manifold topology. We shall be refering to
this foliation as the
\emph{vertical} foliation of the principal bundle and it will be denoted
$P_V$.  Clearly, the
dimension of this foliation is equal to the dimension of the group $G$,
the integrable subbundle of $TP$ being in this case the vertical
subbundle. The codimension is equal to the dimension of the
base space $M$.

Now if in addition a flat connection is given on our principal bundle,
we have another foliation of the total space which we
shall be referring to as the \emph{horizontal} or \emph{flat}
foliation and it will be denoted $P_H$.. We shall study this foliation
extensively in the following subsection. The dimension
of this foliation equals the dimension of the base space and the
codimension equals the dimension of the group. From this one can see
that the vertical and the horizontal foliations of a principal bundle
are \emph{transverse} to each other.

Now the vertical foliation behaves very well; everything is compact
and Hausdorff, as were the spaces we started with to build our
bundle. In this case the general theory of foliations gives
nothing more than the well-known theory of principal bundles. However, the
horizontal foliation can suffer
from various "pathological" deffects and for this reason it is interesting
from the ncg point of view. Let us study it in greater detail.

\subsection{Flat foliation of a principal bundle}

 To begin with, a flat connection on a principal bundle $P$ with
structure group $G$ and base space $M$, corresponds to
reduction of the structure group from $G$ to a subgroup isomorphic to
a normal subgroup of the fundamental group of
the base space $\pi_1(M)$. Moreover a (gauge equivalence class of a) flat conne
ction
also defines a (conjugacy class of a) representation

$$H:\pi _1(M)\rightarrow G$$

If we identify the fundamental group with the group of covering
translations of the universal covering $\tilde{M}$ of $M$ we get an action
$\varkappa$ of $\pi _1(M)$ on $\tilde{M}\times G$  defined as follows:

$$\varkappa :\pi _1(M)\times (\tilde{M}\times G)\rightarrow
(\tilde{M}\times G)$$

$$(\gamma ,(\tilde{m},g))\mapsto (\gamma (\tilde{m}),H(\gamma )(g))$$
where we use the obvious notation $\gamma\in\pi _1(M), g\in G, \tilde{m}\in M$.
This action gives a commutative diagram:

\begin{equation}
\begin{CD}
\tilde{M}\times G@>pr>>\tilde{M}\\
@V\pi VV     @VVqV\\
P'=(\tilde{M}\times G)/\varkappa @>>p>M\\
\end{CD}
\end{equation}
where $pr$ is the canonical projection, $\pi $ is the quotient map by
$\varkappa$, $p$ is uniquely induced by $pr$ and $q$ is just the map from the universal
covering space to the original space.

This construction is called \emph{suspension of the representation
$H$}. One can prove that the map $\pi $ is a covering map and that if
$\digamma :=ImH$ is endowed with the induced topology, then $\xi _H=(P',p,M)$ is
a fiber bundle with fiber $G$, total space $P'$, base $M$, projection $p$ and
structure group $\digamma $.

 To study the geometric properties of suspensions we introduce a new
topology on the total space $P'$ of $\xi _H$. We denote by $G^{\delta }$
the set $G$ supplied with the \textit{discrete} topology. Then the action $\varkappa$ of
$\pi _1(M)$ on $\tilde{M}\times G^{\delta }$ remains continuous and the map
$\pi :\tilde{M}\times G^{\delta }\rightarrow P'$ induces on $P'$ a new topology
which is \textbf{finer} than its manifold topology. We denote by $P^{\delta }$
the set $P'$ supplied with this topology. The topology on $\tilde{M}\times
G^{\delta }$ and the topology $P^{\delta }$ are called the \emph{leaf
topologies}. Then the suspension diagram below is a commutative
diagram of covering maps:
\begin{equation}
\begin{CD}
\tilde{M}\times G^{\delta }@>pr>>\tilde{M}\\
@V\pi VV    @VVqV\\
P^{\delta }@>>p>M\\
\end{CD}
\end{equation}

 The topological space $P^{\delta }$ is not connected unless the fiber
is contractible. A connected component of $P^{\delta }$ is called a
leaf of $\xi _H$. Each point $x=\pi (\tilde{m},g)\in P'$ belongs to exactly
one leaf which is denoted $L_x$ and equals $\pi (\tilde{M}\times {g})$. The
leaves are injectively immersed submanifolds of $P'$ but in general not
embedded. They are transverse to the fibers of $\xi _H$. Conjugate
representations $H$ and $H'$ give suspension bundles $\xi _H$ and $\xi
_{H'}$ which are isomorphic.

 Let now $x=\pi (\tilde{m},g)$. Then the representation 

$$H_x :\pi _1(L_x)\rightarrow G$$
with image $\digamma _g$, is called the \emph{holonomy} representation of
the leaf $L_x$ at the point $x$. The group $\digamma _g$ is the
\emph{holonomy group} of the leaf $L_x$ at the point $x$. $\digamma
_g$ is the isotropy group of $\digamma $ in $g\in G$. Moreover $\pi
_1(L_x)$ is isomorphic to the isotropy group of $\pi _1(M)$ in the
point $g\in G$, namely $\pi _1(L_x)\cong \{\gamma\in\pi _1(M)|H(\gamma
)g=g\}$. See also \cite{hirsch}.\\

There is a topological way to characterise these flat bundles which is by
using  \emph{classifying spaces
for flat bundles} in a fashion analogous for ordinary bundles, namely:\\

 Let $G$ be a connected Lie group and let $G^{\delta }$ denote the
same group with the discrete topology. The \textit{Milnor join construction}
for $G$ defines a connected space $BG$ which is the classifying space for principal
G-bundles. The same construction applied to $G^{\delta }$ yields a
connected topological space $BG^{\delta }$ which is an
\emph{Eilenberg-Maclane space} $K(G,1)$, namely $\pi _1(BG)=G$ and
$\pi _j(BG)=0$ for $j>1$. The inclusion $i:G^{\delta }\rightarrow G$
induces a continuous map $Bi:BG^{\delta }\rightarrow BG$. As sets
these two spaces are the same with the source having \emph{finer}
topology than the range. The difference in these two topologies is
measured by introducing the homotopy fiber $BG'$. This is defined by
first replacing $Bi$ with a homotopy equivalent weak fibration over
$BG$, then take for $BG'$ the (homotopy class of the) fiber. The
description then is just the construction of the \emph{Puppe Sequence}
for $Bi$ (cf \cite{jcw}).

Choose a base point  in $BG^{\delta }$ and consider its image in
$BG$. Then let $\Omega (BG)$ and $P(BG)$ denote the space of based
loops and paths with initial point of $BG$ respectively. Let $e$ be
the end point map of a path. Then one has a fibration
$$\Omega (BG)\hookrightarrow P(BG)\rightarrow BG$$ 
where the second map is $e$. Then define $BG'$ via the homotopy
pull-back diagram:
\begin{equation}
\begin{CD}
\Omega (BG)@>>>\Omega (BG)\\ 
@VVV           @VVV\\
BG'@>>>P(BG)\\
@VVV   @VVeV\\
BG^{\delta }@>>Bi>BG\\
\end{CD}
\end{equation}
 A principal G-bundle $P$ over a manifold $M$ is equivalent to giving
an open covering for $M$ and the transition functions. This data
defines a continuous map $g_P:M\rightarrow BG$. If the transition
functions are locally constant, namely if the bundle $P$ is flat, then $g_P$ can be factored through
$BG^{\delta }$ as a continuous map.

A choice of transition functions which are locally constant is
equivalent to specifying a flat G-structure on $P$. Hence $P$ has a
horizontal foliation whose holonomy map $a:\pi _1(M)\rightarrow G$
defines the classifying map $Ba:M\rightarrow BG^{\delta
}$. Conversely, given a continuous map $Ba:M\rightarrow BG^{\delta }$,
there is induced a representation $a:\pi _1(M)\rightarrow G$ and a
corresponding flat principal G-bundle $P_a=\tilde{M}\times _{\pi _1(M)}G$,
where $\tilde{M}$ is the universal covering of $M$. The topological type of
the G-bundle $P_a$ is determined by the composition
$$g_a:M\rightarrow BG^{\delta }\rightarrow BG$$
The principal bundle is trivial iff $g_a$ is homotopic to the constant
map $M\rightarrow pt$. The choice of the homotopy is equivalent to
specifying a global section on $P_a$.

\subsection{Groupoids and $C^{*}$-algebras associated to Foliations}

 The next step is to associate the \emph{holonomy groupoid}  to any
foliation. In general a groupoid is roughly speaking a small category with
inverses, or more precisely

\textbf{Definition 1:}

A groupoid consists of a set $\Gamma $, a distinguished subset $\Gamma
^{(0)}$ of $\Gamma $,
two maps $r,s:\Gamma \rightarrow \Gamma ^{(0)}$ and a law of composition

$$\circ :\Gamma ^{(2)}:={(\gamma _1,\gamma _2)\in \Gamma \times
\Gamma ; s(\gamma
_1)=r(\gamma _2)}\rightarrow \Gamma $$

such that:\\

1.$s(\gamma _1\circ \gamma _2)=s(\gamma _2)$, $r(\gamma _1\circ
\gamma _2)=r(\gamma _1)$   $\forall (\gamma _1,\gamma _2)\in \Gamma ^{(2)}$

2. $s(x)=r(x)=x \forall x\in \Gamma ^{(0)}$

3.$\gamma \circ s(\gamma )=\gamma$, $r(\gamma )\circ \gamma =\gamma
\forall \gamma \in \Gamma $

4. $(\gamma _1\circ\gamma _2)\circ\gamma _3=\gamma _1\circ (\gamma
_2\circ\gamma _3)$

5. Each $\gamma $ has a two sided inverse $\gamma ^{-1}$, with
$\gamma\gamma ^{-1}=r(\gamma )$ and $\gamma ^{-1}\gamma =s(\gamma )$\\

The maps $r$, $s$ are called \emph{range} and \emph{source} maps.\\

In the category theory terminology, $\Gamma ^{(0)}$ is the space of objects
and $\Gamma ^{(2)}$ is the space of morphisms.

\textbf{Definition 2:}

A \emph{smooth} groupoid $\Gamma $ is a groupoid together with a
differentiable structure on $\Gamma $ and $\Gamma ^{(0)}$ such that the maps $r$, $s$
are submersions and the object inclusion map $\Gamma ^{(0)}\rightarrow
\Gamma $ is
smooth, as is the composition map $\Gamma ^{(2)}\rightarrow \Gamma $.\\ 

 The notion of a $\frac{1}{2}$-\emph{density} on a smooth manifold
allows one to define in a canonical manner the \emph{convolution algebra}
of
a smooth groupoid $\Gamma $.\\

 Specifically, given $\Gamma $, let $\Omega
^{1/2}$ be the
line bundle over $\Gamma $ whose fiber $\Omega _{\gamma }^{1/2}$ at $\gamma\in
\Gamma , r(\gamma )=x, s(\gamma )=y$, is the linear space of maps

$$\rho :\wedge ^{k}T_{\gamma }(\Gamma ^x)\otimes \wedge ^{k}T_{\gamma
}(\Gamma _y)\rightarrow \bf{C}$$
such that 

$$\rho (\lambda\nu )=|\lambda |^{1/2}\rho
(\nu)\forall\lambda\in \bf{R}$$

Here $\Gamma _y={\gamma\in \Gamma ;s(\gamma )=y}, \Gamma ^x={\gamma\in
\Gamma ;r(\gamma )=x}$,
and $k=dimT_{\gamma }(\Gamma ^x)=dimT_{\gamma }(\Gamma _y)$ are the dimensions of
the fibers of the submersions $r:\Gamma \rightarrow \Gamma ^{(0)}$ and
$s:\Gamma \rightarrow \Gamma ^{(0)}$.\\

Then we endow the linear space $C_c^{\infty}(\Gamma ,\Omega ^{1/2})$ of
smooth compactly supported sections of $\Omega ^{1/2}$ with the
convolution product

$$(a*b)(\gamma )=\int _{\gamma _1\circ\gamma _2=\gamma }a(\gamma
_1)b(\gamma _2)$$ 
$\forall a,b\in C_c^{\infty }(\Gamma ,\Omega ^{1/2})$ where the integral on the RHS makes sense since it is the integral of
a 1-density, namely $a(\gamma _1)b(\gamma _1^{-1}\gamma )$, on the
manifold $\Gamma ^x$, $x=r(\gamma )$.\\

One then can prove that if $\Gamma $ is a smooth groupoid and $C_c^{\infty
}(\Gamma ,\Omega ^{1/2})$ is the convolution algebra of smooth compactly
supported $\frac{1}{2}$-densities with involution *, $f^*(\gamma
)=\overline{f(\gamma ^{-1})}$. Then for each $x\in \Gamma ^{(0)}$, the
following defines an involutive representation $\pi _x$ of
$C_c^{\infty }(\Gamma ,\Omega ^{1/2})$ in the Hilbert space
$L^2(\Gamma _x)$:

$$(\pi _x(f)\xi )(\gamma )=\int f(\gamma _1)\xi (\gamma _1^{-1}\gamma
)$$
$\forall\gamma\in \Gamma _x, \xi\in L^2(\Gamma _x)$
\\
The completion of $C_c^{\infty }(\Gamma ,\Omega ^{1/2})$ for the norm
$||f||=sup_{x\in \Gamma ^{(0)}}||\pi _x(f)||$ is a $C^*$-algebra denoted
$C^*_r(\Gamma )$.

Moreover one defines the $C^*$-algebra $C^*(\Gamma )$ as the completion of
the involutive algebra $C_c^{\infty }(\Gamma ,\Omega ^{1/2})$ for the norm 
\\
\\
$||f||_{max}=sup||\pi (f)||$; \{$\pi $ involutive Hilbert space
representation of $C_c^{\infty }(\Gamma ,\Omega ^{1/2})$\}
\\
\\

 After this general introduction to groupoids and to $C^*$-algebras
associated to them, now we pass to
groupoids and $C^*$-algebras associated to foliations.\\

 Let $(V,F)$ be a foliated manifold of codim $q$. Given any $x\in V$ and a
small enough open set $W$ in $V$ containing $x$, the restriction of the
foliation $F$ to $W$ has as its leaf space an open set of $\bf{R}^q$ which we
shall call a transverse neighborhood of $x$. In other words, this open
set $W/F$ is the set of plaques around $x$. Now given a leaf $L$ of $(V,F)$
and two points $x, y \in L$, any simple path $\gamma $ from $x$ to $y$ on $L$
uniquely determines a germ $h(\gamma )$ of a diffeomorphism from a
transverse neighborhood of $x$ to one of $y$. This depends only on the
homotopy class of $\gamma $ and is called the \emph{holonomy} of the
the path $\gamma $. The \textit{holonomy groupoid of a leaf} $L$ is the
quotient of its fundamental groupoid by the equivalence relation which
identifies two paths $\gamma _1$, $\gamma _2$ from $x$ to $y$ both in $L$
iff $h(\gamma _1)=h(\gamma _2)$. Here by the fundamental groupoid of a
leaf we mean the groupoid $\Gamma =L\times L$, $r$, $s$ are the two projections,
$\Gamma ^{(0)}=L$ and the composition is $(x,y)\circ (y,z)=(x,z)$. (From
this one can see that every space is a groupoid). The
holonomy covering $\tilde{L}$ of a leaf $L$ is the covering of $L$ associated to
the normal subgroup of its fundamental group $\pi _1(L)$ given by
paths with trivial holonomy. The \textbf{holonomy groupoid} or \textbf{graph of the
foliation} is the union $\Gamma $ of the holonomy groupoids of its
leaves. Given an element $\gamma $ of $\Gamma $ we denote by $s(\gamma )=x$
the origin of the path $\gamma $ and by $r(\gamma )=y$ its end point,
where $r$, $s$ are the range and source maps as in the general case.

 An element of $\Gamma $ is thus given by two points $x=s(\gamma )$ and
$y=r(\gamma )$ of $V$ together with an equivalence class of smooth paths
: the $\gamma (t)$, $t\in [0,1]$ with $\gamma (0)=x$ and $\gamma
(1)=y$, tangent to the bundle $F$, namely with $\frac{d\gamma }{dt}\in
F_{\gamma (t)}\forall t\in \bf{R}$, identifying $\gamma _1$ and $\gamma _2$
as equivalent iff the holonomy of the path $\gamma _2\gamma _1^{-1}$
at the point $x$ is the identity. The graph $\Gamma $ has an obvious composition
law. For $\gamma _1$ and $\gamma _2$ in $\Gamma $, the composition $\gamma
_1\circ\gamma _2$ makes sense if $s(\gamma _1)=r(\gamma _2)$. The
groupoid $\Gamma $ is by construction a (not necessarily Hausdorff) manifold
of dimension $dim\Gamma =dimV+dimF$.

 \textbf{Definition 3:} The \textbf{$C^*$-algebra of the foliation} is
exactly the $C^*$-algebra of
its graph, as described for arbitrary groupoids above.\\

For our foliations of interest, the graph $\Gamma $ is the following: for the ve
rtical foliation is
just the manifold $P\times G$ whereas for the horizontal
foliation is $P\times _a\tilde{M}$, where $a$ is the representation
from $\pi _1(M)$ to $G$ induced by the flat connection 1-form  (via
the holonomy). Moreover the distinguished subset $\Gamma ^{(0)}$ in both
cases is the manifold we want to foliate, namely $P$, the total space
of our bundle in our case.\\

The $C^*$-algebras associated to our foliations are: for the vertical
foliation is $C(M)$ tensored with compact operators which act as
smoothing kernels along the leaves which in turn is strongly Morita
equivalent to just $C(M)$,
whereas for the horizontal foliation is strongly Morita
equivalent (abreviated to SME) to $C(P)\rtimes \pi _1(M)$. (Note: the
representation of
the fundamental group of the base onto the structure Lie group induced by the flat
connection 1-form used enters the definition of the crossed
product). The first algebra is \textbf{commutative} (up to SME), but the second
\textbf{is not!} It is
for this reason that we can see now that ncg has an important role to
play, in fact we are deeply in the ncg setting. Obviously if the space is
simply connected, i.e. $\pi _1$ vanishes, non-commutativity is lost. We
would like to emphasise that in all cases in the literature where some 
"non-commutative" algebras were used, especially in connection to the
well-known Connes-Lott model for electroweak theory (or even QCD), these
algebras are in fact SME to commutative ones. Hence in terms of topology,
this is not a real non-commutative case.

\subsection{K-classes associated to foliations}

  We shall give the general construction for an arbitrary foliation.\\

Let $(V,F)$ be a foliated manifold and $t=TV/F$ the transverse bundle of
the foliation. The holonomy groupoid $\Gamma $ of $(V,F)$ acts in a natural way
on $t$ by the differential of the holonomy, thus for every
$\gamma\in\Gamma$, $\gamma
:x\rightarrow y$ determines a linear map $h(\gamma ):t_x\rightarrow
t_y$. We denote this action by
$h$. It is not in general possible to
find a Euclidean metric on $t$ which is invariant under the above action
of $\Gamma $. Let $g$ be an arbitrary smooth Euclidean metric on the real vector
bundle $t$. Thus for $\xi\in t_x$ we let $||\xi ||_{g}=(\langle\xi ,\xi
\rangle _{g})^{1/2}$ be the corresponding norms and inner products and drop
subscript $g$ henceforth. Using $g$ we define a $C^*$-module $E$ on the
$C^*$-algebra $C^*_r(V,F)$ of the foliation. Recall that $C^*_r(V,F)$
is the completion of the convolution algebra $C_c^{\infty }(\Gamma ,\Omega
^{1/2})$ which acts by right convolution on the linear space
$C_c^{\infty }(\Gamma ,\Omega ^{1/2}\otimes r^*(t_{\bf{C}}))$ denoted
$\Lambda $ for simplicity and $t_{\bf C}$ is the complexification of the
transverse bundle $t$:

$$(\xi f)(\gamma )=\int _{\Gamma ^y}\xi (\gamma _1)f(\gamma _1^{-1}\gamma
)$$
where $y=r(\gamma )$. Endowing the complexified bundle $t_{\bf{C}}$ with the
inner product associated to $g$ and anti-linear in the first variable,
the following formula defines a $C_c^{\infty }(\Gamma ,\Omega
^{1/2})$-valued inner product

$$\langle\xi ,n\rangle (\gamma )=\int _{\Gamma ^y}\langle\xi (\gamma _1^{-1}),n(\gamma
_1^{-1}\gamma )\rangle $$
for any $\xi ,n\in C_c^{\infty }(\Gamma ,\Omega ^{1/2}\otimes
r^*(t_{\bf{C}}))$
One then checks that the completion $E$ of the space $C_c^{\infty
}(\Gamma ,\Omega ^{1/2}\otimes r^*(t_{\bf{C}}))$ for the norm 

$$||\xi ||=(||\langle\xi ,\xi \rangle ||_{C^*_r(V,F)})^{1/2}$$
becomes a $C^*$-module over $C^*_r(V,F)$. If one takes also the action
$h$
of $\Gamma $ on $t$ into account, with some extra effort one can make $E$ into 
a
$(\Lambda ,\Sigma )$-bimodule (for the definition of the algebra $\Sigma
$ see below). The first construction thus gives us an element $E$ of
$K_0(C^*_r(V,F))$ whereas the second gives $E$ as an element of
$KK_0(\Lambda ,\Sigma )$, \emph{Kasparov's bivariant K-Theory}. (Recall
that the 0th
Kasparov's bivariant K-group in this case consists of stable
isomorphism classes of $(\Lambda ,\Sigma )$-bimodules). We shall use this
action $h$ to define a left action of
$C_c^{\infty }(\Gamma ,\Omega ^{1/2})$ on $E$ by:

$$(f\xi )(\gamma )=\int _{\Gamma ^y}f(\gamma _1)h(\gamma _1)\xi (\gamma
_1^{-1}\gamma )$$

$\forall f\in C_c^{\infty }(\Gamma ,\Omega ^{1/2}), \xi\in E$

One then can prove that for any $f \in C_c^{\infty }(\Gamma ,\Omega ^{1/2})$
the above formula defines an endomorphism $\lambda (f)$ of the
$C^*$-module $E$ whose adjoint $\lambda (f)^*$ is given by

$$(\lambda (f)^*\xi )(\gamma )=\int _{\Gamma ^y}f^{\#}(\gamma _1)h(\gamma
_1)\xi (\gamma _1^{-1}\gamma )$$
\\
where
\\
$$f^{\#}(\gamma ):=\widetilde{f}(\gamma ^{-1})\Delta (\gamma )$$ 
\\
and
\\
$$\Delta (\gamma )=(h(\gamma )^{-1})^{t}h(\gamma )^{-1}\in End(t_{\mathbf{C}}(r(\gamma )))$$

 This shows that unless the metric on $t$ is $\Gamma $-invariant, the
representation $\lambda $ is not a *-representation, the subtle difference
between $\lambda (f)^*$ and $\lambda (f^*)$ being measured by $\Delta
$. In particular $\lambda $ is not in general bounded for the
$C^*$-algebra norms on both $End_{C^*_r(V,F)}E$ and $C^*_r(V,F)\supset
C_c^{\infty }(\Gamma ,\Omega ^{1/2})$. However $\lambda $ is a closable
homomorphism of $C^*$-algebras, namely, the closure of the
graph of $\lambda $ is the graph of a densely defined
homomorphism. Then with the graph norm 

$$||x||_{\lambda }=||x||+||\lambda(x)||$$ 
the domain $\Sigma $ of the closure $\widetilde{\lambda }$ of
$\lambda $ is
a Banach algebra which is dense in the $C^*$-algebra
$\Lambda =C^*_r(V,F)$. The $C^*$-module $E$ is then a
$(\Lambda ,\Sigma )$-bimodule. This
particular module $E$ we constructed here will be the one of the two
main ingredients which define the invariant we want and we shall denote
it $E(F)$.

\subsection{Cyclic classes associated to foliations (transverse
fundamental cyclic cocycle)}

 We begin with some definitions from cyclic homology:

\textbf{Definition 1:}

1. A \emph{cycle} of dimension $n$ is a triple
$(\Omega ,d,\int )$ where $\Omega =\oplus _{j=0}^n \Omega ^j$ is a
graded algebra over ${\bf C}$, $d$ is a differential on $\Omega $'s and
$\int
:\Omega ^n\rightarrow C$ is a closed graded trace on $\Omega $.

2. Let $A$ be an algebra over $\bf{C}$. Then a cycle over $A$
is
given by
a cycle
$(\Omega ,d,\int )$ and a homomorphism $\rho :A \rightarrow \Omega
^0$.

A cycle over $A$ of dimension $n$ is essentially determined by its
\emph{character} which is the following $(n+1)$-linear functional on
$A$:

$$\tau (a^0,...,a^n)=\int \rho (a^0)d(\rho (a^1))d(\rho
(a^2))...d(\rho (a^n))$$
\\
$\forall a^j\in A$
\\
One can then prove that this is a \emph{cyclic cocycle} of $A$,
namely         
it defines a cohomology class in the cyclic homology of $A$ and that
the 
above is a necessary and sufficient statement.

 We shall now describe the transverse fundamental class associated to
foliations. There is a general construction for arbitrary foliations which
is quite involving since one has to \emph{complete} the graded algebra.
This is so because the transverse bundle of the foliation may not be
integrable and in this case derivation along transverse directions will
not be a differential.

 We, however, are primarily interested in our two special kinds of
foliations, the vertical and the horizontal foliation of a principal
bundle. These foliations are transverse, both are integrable so
derivatives are differentials and hence one does not have to complete the
graded algebras. We refer to \cite{conness} for the general construction.
Here we shall only describe the classes wich are associated to our two
foliations:

 The vertical and the horizontal (or flat)
foliations of the total space of our principal bundle $P$ will be   
denoted $(P_V)$ and $(P_H)$ respectively.  One then has that there is a natural
 cycle
for the algebra of each foliation, namely:   

\textbf{Vertical Foliation}, cycle denoted $[P_V]$:

The natural cycle canonically associated to the algebra $C_c^{\infty
}(P\times G,\Omega ^{1/2})$ of the vertical foliation consists of:

1. The graded algebra $C_c^{\infty }(P\times G,\Omega ^{1/2}\otimes 
r^*(\wedge P_H^*))$ where $P_H$ is the \emph{horizontal} subbundle (i.e.
the
transverse bundle to the vertical foliation).

2. The differential $d=d_V+d_H+\theta $ where
$$d_H:C^{\infty }(P,\wedge ^r P_V^*\otimes\wedge ^s P_H^*)\rightarrow
C^{\infty }(P,\wedge ^r P_V^*\otimes\wedge ^{s+1}P_H^*)$$

$$d_V:C^{\infty }(P,\wedge ^rP_V^*\otimes\wedge ^s P_H^*)\rightarrow
C^{\infty }(P,\wedge ^{r+1}P_V^*\otimes\wedge ^s P_H^*)$$
$\theta $ means  contraction with the section $\theta\in C^{\infty
}(P,P_V\otimes\wedge ^2P_H^*)$, where $\theta $ is defined by:

$\theta (p_H(X),p_H(Y)=p_V([X,Y])$
for any pair of horizontal vector fields $X, Y \in C^{\infty }(P,P_H)$
and $(p_H,p_V)$ is the isomorphism $TP\rightarrow P_H\oplus P_V$ given by
$P_H$.

3.The trace is defined via

$$\tau (w)=\int _{\Gamma ^{(0)}} w $$
where $\Gamma $ is the graph of the vertical foliation $\Gamma =P\times
G$.\\

\textit{Similarly one defines a fundamental class for the horizontal
foliation}.

One then can define the character of these cycles---essentially the
trace---which is a
class in the cyclic homology of  the appropriate algebra for each
foliation \cite{connes2}.

\textbf{Note:}

Since now we have a cyclic homology class, say $\phi $ of the algebra of
the
foliation, say $\Lambda $, we automatically have a map

$$ K_i(\Lambda )\rightarrow \bf{C}$$
given by  pairing it with K-group elements ($i=0,1$ above) to get index
theorems
for \emph{leafwise} elliptic operators. Let us mention here that the
analytic
\emph{Index}
of an operator \emph{elliptic along the leaves of an arbitrary foliation}
say
$(V,F)$, is an element of $K_0(C^*(V,F))$, being in fact a
generalisation of the index of families of elliptic operators
considered by Atiyah and Singer. (In the Atiyah-Singer case of
families of elliptic operators one is
dealing with the foliation induced by the fibration, which is the
commutative geometry case). The operator \emph{itself} which is elliptic
along the leaves of the foliation is an element of $KK(C^*(V,F),C(V))$.\\

\section{Invariant for the nl$\sigma $m}

 The final step then is to make use of the general formula for pairings
between K-homology and cyclic homology. In
more concrete terms, one has:

\textbf{Definition:}

Let $A$ be an algebra. Then the following equality defines a
bilinear
pairing between K-theory and
cyclic homology:

1. Even case: $K_0(A)$ and $HC^{ev}(A)$:

$$\langle [e],[\phi ]\rangle :=(m!)^{-1}(\phi \# Tr)(e,...,e)$$
\\                                                     
for $e\in K_0(A)$ using the idempotents' description and $\phi\in
HC^{2m}(A)$ and where $\#$ is the \emph{cup product} in cyclic   
homology (see for instance \cite{connes3} for the precise definition).

2. Odd case:  $K_1(A)$ and $HC^{odd}(A)$:

$$\langle [u],[\phi ]\rangle =\frac{1}{\sqrt{2i}}2^{-n}\Gamma
(\frac{n}{2}+1)^{-1}(\phi \# Tr)(u^{-1}-1,u-1,u^{-1}-1,...,u-1)$$
\\

This is an important point because by pairing
the $C^*$-module $E$ we constructed previously naturally associated to the foli
ation considered with
the cyclic cocycle naturally associated to the foliation, we get an
invariant for arbitrary foliations. In particular  if we apply
this to the horizontal foliation, we get a \emph{complex number} which is
an
 invariant for the nl$\sigma$m. Namely one has:
$$\langle [E(P_H)],[P_H]\rangle
=(m!)^{-1}((P_H)\#Tr)(E(P_H),...,E(P_H))\in \bf{C}$$
In more concrete terms, assuming that $E(P_H)\in M_k(C^{*}(P_H))$ for some
$k$ (where $C^{*}(P_H)$ is the corresponding $C^{*}$-algebra to the
horizontal
foliation) and $[P_H]\in Z^{q}(C^{*}(P_H))$ where $Z$ denotes cyclic
cocycles and $q$ is the codimension of the horizontal foliation, then
$(P_H)\#Tr\in Z^{q}(M_k(C^{*}(P_H)))$ is defined by
$$((P_H)\#Tr)(a^0\otimes m^0,...,a^q\otimes
m^q)=(P_H)(a^0,...,a^q)Tr(m^0...m^q)$$ 
for any $a^i\in C^{*}(P_H), m^i\in M_k({\bf C})$, $i=1,...,q$.\\

\emph{Note:}
 Let us mention that the odd case formula is related to the $\eta $
invariant for leafwise elliptic operators, see \cite{douglas1},
\cite{douglas2} which in turn is related to global anomalies and to the 
Freedman-Townsend invariance (cf
\cite{witten2}, \cite{Z2}, \cite{townsend}, \cite{henneaux}).

\section{An example: principal fibre bundles}

In order to get some more insight to this pairing we shall try to
calculate it for the case of principal
bundles (vertical foliation) which is the simplest example. 

We begin by describing the graph in detail:
the set $\Gamma $ in this case is the manifold $P\times G$, the
distinguished subset $\Gamma ^{(0)}=P\times \{e\}$ and denoting the
action (on the right) of $g\ni G$ on $p\ni P$ simply by $(p,g)\mapsto
pg$, one has that the range and source maps are respectively
$r(p,g)=p$ and $s(p,g)=pg$, the inverse $(p,g)^{-1}=(pg,g^{-1})$ and
the law of composition is $(p_{1},g_{1})\circ
(p_{2},g_{2})=(p_{1},g_{1}g_{2})$ if $p_{1}g_{1}=p_{2}$. Obviously the
set $\Gamma ^{(2)}=P\times G\times G$.
 
Moreover we recall that the $C^*$-algebra for the vertical foliation
is strongly Morita equivalent to $C(M)$,

We now make use of two important facts:

1. $K_0(C(M))=K^0(M)$, namely \emph{Serre-Swan theorem}\\

and\\

2. $H^{*}_{cont}(C(M))=H_{*}(M)$\\

where on the RHS we have the ordinary homology of $M$ (with complex
coefficients) and by definition for the LHS we have
$H^{*}(C(M)):=Lim_{\rightarrow} (HC^{n}(C(M)),S)$
(see \cite{connes3} for explanations of the notation), $HC^{*}$
denotes cyclic homology and "cont" means
restriction to continuous linear functionals.

The first fact says that for commutative $C^*$-algebras one gets
Atiyah's topological K-theory for the underlying space (described in
terms of stable isomorphism classes of complex vector bundles over the
space considered) and the second says that in the commutative case
again cyclic homology is "roughly speaking" the ordinary homology of the
underlying space (and thus we see that non-commutative geometry
reduces to ordinary geometry in the commutative case).

 Since we have these two results in our desposal, we shall try to reduce
the whole discussion in terms of bundles and ordinary homology theory
because this is more comprehensible.

In order to describe the pairing then we need the transverse fundamental
cyclic cocycle: we shall give a simple dimensional argument here; the
exact computations are rather too technical to be presented in grater
detail. The cyclic cocycle we will get from the vertical
foliation  will be of dimension equal to the
codimension of the vertical foliation which is equal to the  dimension
of the base space of our bundle. Moreover as we mentioned above, in
this case the $C^*$-algebra of this foliation is SME to the algebra of
functions on the base $C(M)$. This is a commutative $C^*$-algebra whose
cyclic homology is moreorless the de Rham cohomology of the base
space. Hence it is not too hard to suspect that we get a top homology
class, which  in fact turns out to be the
fundamental class of our base space [M].

For the module denoted $E$ above in this case one uses the following fact:
it is a consequence of Serre-Swan theorem mentioned above that the
link between topological K-theory and K-theory of commutative
$C^*$-algebras is that given a complex vector bundle over $M$ (thus a
topological K-class), one considers the corresponding $C(M)$-module of smooth
sections of the given complex vector bundle. Thus in this case the
module $E$ we get is $C^{\infty }_{c}(P\times G,\Omega ^{1/2}\otimes
r^{*}(t_{\bf C}))$, hence we can recover the corresponding complex vector
bundle over $M$ as follows:

If we denote by $(P,\pi ,G,M)$ our original principal bundle and we
consider the vertical foliation, then its normal bundle $t$ would be
$\pi ^{*}(TM)$, where $TM$ is the tangent bundle of $M$. We prefer the
topological K-theory description which in this case is rather easy to
read: the bundle associated to this $C(M)$ module $E$ is:
\begin{equation}
\begin{CD}
\Omega ^{1/2}\otimes pr_{1}^{*}\pi ^{*}(TM)@>>>\pi ^{*}(TM)@>>>TM\\
@VVV     @VVV       @VV\tau _{M}V\\
P\times G@>>pr_{1}>P@>>\pi >M\\
\end{CD}
\end{equation}
where the fibre of the line bundle $\Omega ^{1/2}$ is the linear space
of maps
$\rho :\wedge ^{dimG}T_{\gamma }(G)\otimes \wedge ^{dimG}T_{\gamma
}(G)\rightarrow {\bf C}$ satisfying the well-known property for 1/2-densities.
Hence in this case the result will be the number we get if we take the
bundle $\Omega ^{1/2}\otimes pr_{1}^{*}\pi ^{*}(TM)$ \emph{seen as a
bundle over $M$}, then apply the ordinary Chern character to it and
integrate over $M$. The result will be a combination of the Pontryagin
class of $TM$ and the second Chern class of $P$ (recall that we assumed
$M$ to be 4-dim and $P$ is an SU(2) bundle)
which is something expected.

 There are some subtleties though: we have the bundle $\Omega
^{1/2}\otimes pr_{1}^{*}\pi ^{*}(TM)$ over the graph which is $P\times G$.
We want to see this as a bundle over $M$. We consider firstly the factor
$pr_{1}^{*}\pi ^{*}(TM)$. This is indeed a bundle over $M$ with fibre
$G\times G\times {\bf R}^{dim M}$ but this is neither a vector nor a
principal bundle and in order to talk about characteristic classes one
actually needs the one or the other. In order not to change the topology
then, which is what we are mainly interested in, we can consider the
vector bundle $TM\otimes adP$ instead, where $adP$ is the adjoint bundle
to $P$. To study formally the classes of $TM\otimes adP$ is an exercise in
mathematics (we forget the pull-backs since they can be treated easily).
The point is that we shall get a combination of the Pontryagin classes of
$TM$ (or Chern if we complexify) and of Chern classes of $P$. The later is
known since the bundle is given whereas the former can be computed from
topological information of $M$ itself. For example for simply
connected closed 4-manifolds one has from the Hirzebruch signature
formula that (see \cite{hirzebruch}):
$$p_1 = 3\tau =3(b^{+} - b^{-})$$
where $\tau $ is the signature.

 As about the other factor, the 1/2-densities of the graph, seen as a rank
1 real 
bundle over the
graph $P\times G$, is rather dull. It will be determined by $\omega _1$,
the first Stiefel-Whitney class: either trivial ($\omega _1=0$) or it is
non-orientable ($\omega _1=1$). (\emph{Note:} half densities over a 
complex
manifold say $N$ with $dimN=k$ are slightly more complicated: in this case
its class will be
$\frac{1}{2}c_1(\wedge ^kT^*)$, where 
 $\wedge ^kT^*$ is the canonical bundle, so
it will correspond to $spin ^c$ structures on $N$). But we still have the
same problem that over $M$ it is neither a vector nor a principal bundle.
This can be overcome as before; the point is that since it is a dull
bundle over $P\times G$, the projection does not change anything, so as a
bundle over $M$ it will be determined by the topology of $P$, hence we
also have Chern classes of $P$.

 What we gave above was a qualitative description and the lesson was that
the invariant will be some combination of Chern numbers of $P$ and the
Pontryagin number of the tangent bundle $TM$ of $M$. The key point is
that $TM$ appears because it is the \emph{transverse bundle} of the
vertical
foliation. We expect then that characteristic classes of the transverse
bundle should be important in general.

What about the flat foliation then, which is the case which is related to
nl$\sigma $m in general and to M-theory in particular in physics? This is
a purely nc case. Computations are much
harder and a picture involving bundles is impossible since we do not have
Serre-Swan theorem. Moreover cyclic homology of nc algebras has no
relation to the usual topology. We can still say something though: first
of all, since we have a flat principal bundle, all its characteristic
classes vanish, so we get nothing from them. Since the base space $M$ is
not simply connected, gauge inequivalent classes of flat connections are
characterised by their holonomy. So we expect the holonomy to play some
role.

 Moreover, from the vertical foliation case we saw that the normal bundle
of the foliation also plays a vital role.

\textbf{Note:}
It is not always true that in the commutative case cyclic homology
identifies well with ordinary homology, in fact this is always
essentially true only for $H_0$. There may be complications.
However this point will not be treated in this article with greater
detail.  See \cite{connes5}.

\section{Relation to Physics}

 There are three cases in physics where this invariant may play some role:

\emph{1. Nl$\sigma $m.}

 As it is well-known, $\sigma $ models classically describe harmonic maps
between two Riemannian manifolds (target and source spaces). From the
genral remarks we made in the
preceeding section, we said that we expect the invariant to include the
holonomy of the flat connection (namely $\pi _1$ of the source space) plus
the topology of the space of leaves (target space). Hence this invariant
should contain information about the topology of both manifolds
involved in
$\sigma $ models. Characteristic classes of foliations may provide the way
to calculate the invariant (analogue of Chern-Weil theory, see
\cite{tondeur}).

Moreover, in the flat foliation case the invariant describes the
topological charge of the M-theory Lagrangian density suggested in
\cite{Z1}

Another application is the following (we thank Dr S. T. Tsou for
pointing this out to us): we know from Polyakov
(\cite{polyakov}) that Yang-Mills theories can be formulated as nl$\sigma
$m on the \emph{loop space}. This point of view very recently
exibited some very nice dualities of the Standard Model, see
\cite{tsou}. Hence the invariant, for the flat foliation case (since
nl$\sigma $m can be thought of as a flat bundle with structure group the
isometries of the target space),
maybe of some relevance also for Yang-Mills theories, the setting however 
will involve loop spaces now! We do not know exactly what its physical
significance would be. Moreover doing K-theory on loop spaces (infinite
dimensional manifolds) is considerably harder. There are however some path
integral techniques. (see again \cite{tsou} and references therein).

 \emph{2. Instantons with non simply connected boundary}.

 Following largely the case of instantons we suggest that this invariant is related to
interpolation between \emph{gauge inequivalent vacua} which exist due
to the non simply connectedness of the space considered. Clearly there
is \emph{extra} degeneracy of the vacuum coming from the fact the our
space is
\textsl{not simply connected}. This degeneracy is of different origin than that
of
instantons since as it is well-known for ordinary instantons the
degeneracy comes from the \textsl{different topologies} of the bundle
considered. In more concrete terms we suppose that this invariant will
be relevant in the following case: let us assume that we try to follow
the discussion in the BPST famous paper on instantons \cite{BPST}; if we assume
that we have a space whose boundary is not just $S^3$ as in that case
but a 3-manifold which has a non-trivial $\pi _{1}$. In this case we want
the potential to become flat (pure gauge) on the boundary. However if the
boundary is a 3-manifold with a non trivial fundamental group, then flat
connections are not unique (up to gauge equivalence of course). We know
more specifically that gauge equivalent classes of flat connections are in
1-1 correspondence with  conjugate classes of representations of the
fundamental group onto the structure group considered. Thus in this case
the flat connection we choose will not be unique. This \emph{extra
}degeneracy of
the vacuum comes from the different possible choices of flat connection,
which is something noticed for the first time.
The invariant is related to interpolation between these extra vacua.

We
expect then some relation with the so called \emph{ALE
gravitational instantons} which are important both in quantum gravity
and in gauge theory \cite{barrett}, \cite{kronheimer}. 

\emph{3. Gravity, Non-Commutative Topological Quantum Field Theories
(ncTQFT for brevity).}

 In ordinary Yang-Mills
theory, gauge transformations are described as
automorphisms of the bundle (namely fibre preserving maps) which
\emph{induce the identity on the base space} (cf for example
\cite{donald}). Sometimes these are called
\emph{strong} bundle automorphisms. If one wants to generalise this
picture and attempts to include the symmetry of general relativity, namely
local diffeomorphisms of the base space, then there is a problem because
there are local diffeomprphisms of the base space which can not be induced
by
bundle automorphisms (cf \cite{shard}). In simple words: there are "more" 
local diffeomorphisms of the base space than bundle automorphisms. The way
that theoretical
physicists usually try to go arround this problem is - to begin with,
supersymmetry and finally, -  supergravity. The origins of supersymmetry
are actually quantum mechanical (multiplets with same number of bosons and
fermions plus symmetry between particles of different spin which should
exist, if all interactions are eventually unified since gauge particles
have spin-1 whereas the graviton is supposed to have spin-2. Another
point of view is to examine the largest possible symmetry of the
S-matrix elements in the framework of relativistic quantum field theory
on Minkowski space). To make the long
story short, based on the Coleman-Mandula theorem (which is responsible
for introducing anti-commuting coordinates), what one actually does
(following the superspace formalism for N=1 supersymmetry coming from
observation that Minkowski space is actually Poincare/Lorentz) is to
enlarge
the base manifold (assumed to be spacetime) by adding some fermionic
dimensions (non-commuting coordinates), thus obtaining another space, the 
socalled superspace. This superspace, in an analogous fashion, can be seen
as the quotient space superPoincare/Lorentz. Supersymmetric Yang-Mills
theories are principal G-bundles over superspace with G some compact
and connected Lie group (usually SU(N)) whereas supergravity
can be seen as a principal G-bundle over superspace where G is the
superPoincare group (generalisation of Einstein-Cartan
theory). However the so-called Noether technique which makes local a
rigid supersymmetry and which is used mainly to
construct supersymmetric interacting Lagrangians actually suggests
that supersymmetric Yang-Mills theories can be equivalently be seen as
principal G-bundles over ordinary spacetime with G some Super Lie Group.

 Letting alone some severe criticism of supersymmetric theories (e.g.
positive metric assumption), especially
when the discussion
comes to supergravity-- (the most important experimental problem of
supersymmetric theories is the fact that none of the superparteners of
particles has ever been observed, the way phenomenologists try to
overcome this problem is to assume spontaneously breaking of
supersymmetries; two of
the main theoretical problems are: the aspect of supergravity as a local
gauge theory 
which is not completely mathematically justified, for example in N=8
D=4 supergravity theory coming from D=11 N=1
supergravity which is supposed to be the best candidate for unification
and according to recent progress one of the two low energy limits of
M-theory, local diffeomorphisms are supposed to come from gauging the
group O(8), an assumption which is based on the observation that ordinary
gravity comes from gauging the Poincare group, something which is wrong
because of the existence of the "shouldering" form on arbitrary curved
manifolds; the second important problem is that most of the extended
supersymmetric and supergravity theories are actually up to now
 formulated only "on-shell", namely they are essentially classical
theories - for N=1
supergravity though there is another problem, one has more than one
"off-shell" formulations, this in fact has now an explanation from the
recently observed
string/5-brane duality in D=10 (old brane-scan); - trying to be fair, we must mention that the
good features of such theories are that they offer probably the only
up to now known hope for unification plus the fact that they give "less
divergent" theories, something essential for perturbative quantum fiels
theories),-- we would like to propose here another
approach; our approach is more in the spirit of non-perturbative quantum
field theories, in fact topological quantum field theories: \emph{instead of enlarging the base manifold by considering
anti-commuting coordinates, we chose to relax the
fibre preserving condition}, meaning that now we allow "bundle" maps which 
are not fibre-preserving; in such a case, fibres may be "mixed
up", for example they may be "tilted" or "broken".
The resulting structure after applying these more general
transformations to our original bundle may no longer be a fibre bundle,
but it will still be a foliation. In this
case however what we will get as the quotient space will not necessarily 
be the
manifold we had originally in our bundle construction (supposed to be
space-time) but another space of leaves with the same dimension and
maybe very
different topology. In this case the dimension of the leaves is kept
fixed, equall to the dimension of the Lie algebra considered; had we
changed
that, the dimension of the space of leaves would have changed accordingly.
This picture is quite close to the picture that string theorists
patronise, namely that space-time is not fixed but it emerges as a ground
state from some dynamical process.

In fact there is a deep result due to Thurston which for a given
manifold M, say, it relates the group of local diffeomorphisms of M
with the group of foliations of M \cite{thurston}. In particular
Thurston proves that one has an isomorphism in cohomology after some shift in
the degrees between the classifying space of local diffeomorphisms and
the classifying space of foliations of a closed manifold M. If we take M to be
the total space of a principal bundle P over spacetime, then obviously
local diffeomorphisms of the base space are included to local
diffeomorphisms of the total space which in turn are very closely
related to the group of foliations. Hence at least in principle looking at
the group of foliations of the total space of a principal bundle
provides a framework which is rich enough in order to encorporate
local diffeomorphisms of the base space, something we need in order to
relate general relativity with Yang-Mills theory symmetries and this
framework is mathematically rigorous.

The group then of all foliations of the total space with fixed codimension
is huge. It definitely contains all foliations which are "regular" enough
in order to
get manifolds diffeomorphic to our original one. Yet foliations can be
really nasty: in this case the quotient space may not be a manifold at all
but a "quantum" topological space. All these cases need to be studied. For
the moment we know that whenever the foliations have a corresponding $C^*$
algebra which is SME to a commutative one, then the space of leaves will
be a compact Hausdorff topological space of the same dimension. If 
the $C^*$-algebras of two foliations can be related with a
*-preserving homomorphism, then the corresponding quotient spaces
will be homoeomorphic.
What is the
appropriate condition on the $C^*$ algebras in order to get diffeomorphic
manifolds, we do not know (this point is of particular interest in 4-dim
due to the existence of the so-called "exotic structures" for
4-manifolds). The main point here is that we can "control"
how much non-commutativity we want in the $C^*$- algebra and then see what
this means topologically. At this point we would like to recall that
mathematically, going from classical physics to quantum is going from
commuting algebras to non-commuting ones. The essence of Planck's constant
then
is that it tells us "how much" non-commutativity we want. Moreover there
is the fundamental theorem for $C^*$- algebras representations, namely
that
for each $C^*$- algebra (commutative or not), there exists a Hilbert space
whose space of bounded operators is actually "the same" as the original
$C^*$- algebra. Hence for each foliation there exists a corresponding
$C^*$-
algebra (commutative or not), a corresponding topological space (space of
leaves which may be a manifold or a quantum topological space
respectively) and finally, a Hilbert space as a representation space!

 Let us now turn to something related to the above but a little more
concrete: 
for the moment let us consider the case where the dimensions of the
leaves and of the space of leaves are kept
fixed. This situation has some similarities with quantum gravity seen as a
TQFT (in fact we generalise that picture and we present a way to consider 
unified theories-namely gravity and Yang-Mills theories-as
Non-Commutative Topological Quantum Field Theories).
In \cite{barrett} it was argued that TQFT may provide a framework
which is rich enough for the development of a quantum theory of gravity.
In that aspect, space-time was treated as an unquantized object whereas
the metric was quantum mechanical. The idea in TQFT  framework is to find
an invariant $Z(M)$ for a topological space $M$ and then one seeks for a
Lagrangian density whose partition function yields the invariant $Z(M)$,
see \cite{WIT}. One has to be a little more careful though in order for
the Atiyah's axioms for TQFT to be satisfied \cite{AT}. In quantum
mechanics one usually has a space of quantum states associated to a given
system. Often this space of states refers to a particular instant of time,
which can be represented in a 4-dimensional world by a space-like
hypersurface. In TQFT this vector space appears as part of the definition,
when the space-time $M$ has a boundary, i.e. a space of dimension 1 less.
In more
concrete terms, to a $(d-1)$-dim space $\Sigma $ we associate a vector
space $V(\Sigma )$ and to each $d$-dim space $M$ with $\partial M=\Sigma $
we associate a vector $Z(M)$, the partition function of the space $M$.
This point can be generalised, in fact $\Sigma $ can be any embedded
submanifold of $M$ with dimension 1 less. One interpretation of these
conditions is that $\Sigma $ represents the "present instant" of time and
that the vectors in $V(\Sigma )$ which are determined by various choices
of observables represent a memory of past facts. The primary problem
nonetheless is the construction of invariants for spaces and the state
spaces and partition functions for spaces with boundary are usually
obtained as a by-product. Since by our proposal above  one can
 end up with
quantum topological spaces as spaces of leaves of foliations, one can call
this theory
\emph{non-commutative} topological quantum field theory and we believe
that this can provide a framework for quantum unified theories (including 
Yang-Mills and gravity).

 The picture we have then is the following: we start with a G-bundle $P$
over a 4-manifold $M$. From symmetry
considerations, namely we want to include local diffeomorphisms of the
base space and relate
them to bundle automorphisms (hence relating general relativity to gauge
theory), we end up to consider all dimG-dim foliations of $P$.
Automatically the 4-dim space which is the space of leaves is somehow
"quantized", namely it is forced to have one of the leaf topologies. This
is a difference with TQFT as explained in \cite{barrett} where
the metric was quantised but space-time
was unquantized (needless to say, in such a case as ours, the metric is
quantised
too automatically). Moreover, for each foliation we have a quotient space
of leaves and hence an invariant $Z(M)$
which is a complex number. The boundary of course can be added  with
its
vector space attached to it, one however has to examine what happens to it
as foliations vary. The Lagrangian density whose partition function is
this invariant for foliations is an open question. It should be related to
characteristic classes for foliations. A good indication for that is the
fact that the \emph{$\Gamma _q$ functor} of q-dim Haefliger structures or
$\Gamma _q
$ structures as they are known in topology (and
hence foliations which is an example of a $\Gamma _q$ structure, where $q$
is the codimension of the foliation) \emph{is representable}, see
\cite{james}. So for the moment we do not have a "full" specific ncTQFT
since we have the invariant but not the appropriate Lagrangian density. We
presented though a generalisation of TQFT for non-commutative cases. 

Another possible application might be the ability to construct \emph{deformed}
Yang-Mills theories, see \cite{schwarz}. In that paper, some new
compactifications of the IKKT matrix theory on non-commutative tori were
introduced which, in a certain sense, could be realised as deformed
Yang-Mills theories. Clearly in this case our invariant will be the
"instanton number" of these deformed Yang-Mills theories.

 This picture also
suggests that the above described
non-commutative topological quantum field theories can be seen as emerging
from M-theory compactified down to some non-commutative spaces (tori or
other).

\section{M-Theory}

In this section we shall present an application to M-Theory. Since it
is more extensive, we give it separately.

We know that M-Theory consists of membranes and
5-branes living on an 11-manifold (\cite{duff}, \cite{west}) and it is non perturbative. This theory
has a very intriguing feature: we can only extract information about it
from its limiting theories, namely either from D=11 N=1 supergravity or
from superstrings in D=10. This is so because this theory is
\emph{genuinly} non
perturbative for a reason which lies in the heart of manifold topology:

Let us recall that in string theory, the path integral involves summation
over all topologically distinct diagrams (same for point
particles of
course). Strings are 1-branes hence in time they swep out a 2-manifold. At
the tree level then we need all topologically distinct simply connected
2-manifolds (actually there is only one, as topology tells us) and for
loop corrections, topology again says that topologically distinct non
simply connected 2-manifolds are classified by their genus, so we sum up
over all Riemann surfaces with different genus.

It is clear then that for a perturbative quantum field theory involving
p-branes we have to sum upon all topologically distinct (p+1)-dim
manifolds: simply connected ones for tree level and non simply connected
ones for loop corrections. Thus \emph{we must know before hand the
topological
classification of manifolds} in the dimension of interest. That is the
main problem of manifold topology in mathematics.

But now we face a deep and intractable problem: geometry tells us,
essentially via a no-go theorem which is due to Whitehead from late '40's,
that:
\emph{"we cannot classify non simply connected manifolds with dimension
greater or equal to 4"}! Hence for p-branes with p greater or equal to 3,
all we can do via perturbative methods is up to tree level!

What happens for 3-manifolds then (hence for membranes)?
The answer from mathematics is that we \emph{do not know} if all
3-manifolds can be
classified! So even for 2-branes it is still unclear whether perturbative
methods work (up to all levels of perturbation theory)!\\

The outlet from this situation that we propose here is not merely to look
only at non perturbative aspects of these theories (i.e. the soliton
part of the theory) and then apply S-duality, as was done up to now,
but to abandon perturbative methods completely from the very beginning.
There is only one way known up to now which can achieve this "radical"
solution to our problem: \emph{formulate
the theory as a Topological Quantum Field Theory} and hence get rid of all
perturbations once and for all.\\

Let us explain how this can be acieved.\\

Our approach is based on one physical \emph{"principle"}:\\

\textsl{A theory containing \emph{p-branes} should be formulated on an
m-dim manifold
which \emph{admits $\Gamma _q$-structures}, where $q=m-p-1$}.\\

{\bf N.B.} 

Although we used in our physical principle $\Gamma _q$-structures
which are more general than foliations, we shall use both these terms
meaning essentially the same structure. The interested reader may
refer to \cite{lawson} for example to see the precise definitions which are
quite complicated. The key point however is that the difference
between $\Gamma _q$-structures (or Haefliger structures as they are
most commonly known in topology) and codim-q foliations is essentially the
difference between \textsl{transverse} and \emph{normal}. This does
not affect any of what we have to say, since Bott-Haefliger theory
of characteristic classes is formulated for the most general case, namely
$\Gamma $-structures. We would also like to mention the relation
between $\Gamma $-structures and $\Omega $-spectra which is currently
an active field in topology.\\

(For D-branes we need a variant of the above principle, namely we need
what are called \emph{plane foliations} but we shall not elaborate on this
point here).\\

One way of thinking about this principle is that it is analogous to
the ``past histories'' approach of quantum mechanics. Clearly in
quantum level one should integrate over all foliations of a given codim.\\

\textsl{A piece of warning here:} this principle does not imply that \emph{all}
physical process between branes \emph{are} described by foliations. Although
the group of foliations is huge, in fact comparable in size with the
group of local diffeomorphisms \cite{thurston}, and foliations can be really ``very
nasty'', we would not like to
make such a strong statement. What is definitely true though is that
\emph{some} physical process \emph{are indeed} described by foliations, hence
\textsl{at least} this condition \emph{must} be satisfied because of them.\\

{\bf Note:}

Before going on furter, we would like to make one crucial remark: this
principle puts severe restrictions on the topology that the underlying
manifold may have, in case of M-Theory this is an 11-manifold. It is also
very important if the manifold is \emph{open} or \emph{closed}. This may
be of some help, as we hope, for the compactification problem of string
theory or even M-Theory, namely how we go from D=10 (or D=11) to D=4 which
is our
intuitive dimension
of spacetime. We shall address this question in the next section. The final comment is this: this principle puts
\emph{absolutely no restriction} to the usual quantum field theory for
point particles in D=4, e.g. electroweak theory or QCD. This is so because in
this case spacetime is just ${\bf R^4}$ which is non compact and we have
0-branes (point particles) and consequently 1-dim foliations for which the
integrability condition is trivially satisfied (essentially this is due
to a deep result of Gromov for foliations on open manifolds, which
states that all open manifolds admit codim 1 foliations; in striking contrast,
closed  manifolds admit codim 1 foliations iff their Euler
characteristic is zero, see
for example in \cite{lawson}, \cite{gromov} or references therein).\\

If we believe this principle, then the story goes on as follows: we are on
an
11-manifold, call it M for brevity and we want to describe a theory
containing 5-branes for example (and get membranes from S-duality). Then M
should admit 6-dim foliations or equivalently codim 5 foliations. We know
from Haefliger that the $\Gamma _q $-functor, namely the functor of codim
q Haefliger structures and in particular codim q foliations, is
representable. Practically this means that we can have an analogue of
Chern-Weil theory which characterises foliations of M up to homotopy using
cohomology classes of M. (One brief comment for foliations: one way
of describing Haefliger
structures more generally is to say that they \emph{generalise fibre
bundles in exactly
the same way that fibre
bundles generalise Cartesian product}. This observation is also
important when mentioning \textsl{gerbes} later on).\\

In fact it is proved that the correct cohomology to classify Haefliger
structures up to homotopy (and hence foliations which constitute a
particular example of Haefliger structures) is the \emph{Gelfand-Fuchs}
cohomology. This is a result of Bott and Haefliger, essentially
generalising an earlier result due to Godbillon and Vey which was dealing
only with codim 1 foliations, \cite{gv}.\\

Now we have a happy coincidence: the Bott-Haefliger class for a codim 5
foliation (which, recall, is what we want for 5-branes on an 11-manifold)
is exactly an 11-form, something that fits well with using it as a
Lagrangian density!

The construction for arbitrary codim q foliations goes as follows: let
$F$ be a codim q foliation on an m-manifold M and suppose its
normal
bundle $\nu (F)$ is orientable. Then $F$ is defined by a
global decomposable q-form $\Omega $. Let $\{(U_{i},X_i)\}_{i\in
I}$ be a locally finite cover of distinguished coordinate charts on M with
a smooth partition of unity $\{\rho _i\}$. Then set

$$ \Omega = \sum _{i\in I} \rho _{i}dx_{i}^{m-q+1}\wedge ... \wedge
dx_{i}^{m} $$

Since $\Omega $ is integrable,

\begin{equation}
d\Omega =\theta \wedge \Omega ,
\label{eq1}
\end{equation}

where $\theta $ \emph{some} 1-form on M. The (2q+1)-form

\begin{equation}
\gamma = \theta \wedge (d\theta )^{q},
\label{eq2}
\end{equation}

is closed and its de Rham cohomology class is independent of all choices
involved in defining it, depending only on homotopy type of $F$.
That's the class we want.

Clearly for our case we are on an 11-manifold dealing with 5-branes, hence
6-dim foliations, hence codim 5 and thus the class $\gamma $ is an
11-form.

This construction can be generalised to arbitrary $\Gamma
^{r}_{q}$-structures as a mixed de Rham-Cech cohomology class and thus
gives an element in $H^{2q+1}(B\Gamma ^{r}_{q};{\bf R})$, where $B\Gamma
^{r}_{q}$ is the classifying space for $\Gamma ^{r}_{q}$-structures.
Note that in fact the BHGV class is a cobordism invariant of codim q foliations
of compact (2q+1)-dim manifolds. This construction gives \emph{one}
computable characteristic class for foliations. Optimally we would
like a generalisation of the Chern-Weil construction for $GL_q$. That
is we would like an abstract GDA with the property that for any codim
q foliation $F$ on a manifold M there is a GDA homomorphism
into the de Rham algebra on M, defined in terms of $F$ such
that the induced map on cohomology factors through a universal map
into $H^*(B\Gamma _{q}^{r};{\bf R})$. This algebra is nothing more
than the \emph{Gelfand-Fuchs} Lie coalgebra of \emph{formal} vector fields in
one variable. 

More concretely, let $\Gamma $ be a transitive Lie-pseudogroup acting
on ${\bf R^n}$ and let $a(\Gamma )$ denote the \textsl{Lie algebra of
formal $\Gamma $ vector fields} associated to $\Gamma $. Here a vector
field defined on on $U\subset {\bf R^n}$ is called a $\Gamma $ vector
field if the local 1-parameter group which it engenders is $\Gamma
$ and $a(\Gamma )$ is defined as the inverse limit
$$a(\Gamma )=lim_{\leftarrow}a^{k}(\Gamma )$$
of the k-jets at 0 of $\Gamma $ vector fields. In the pseudogroup
$\Gamma $ let $\Gamma _0 $ be the set of elements of $\Gamma $ keeping
0 fixed and set $\Gamma _{0}^{k} $ equal to the k-jets of elements in
$\Gamma _0 $.\\

Then the $\Gamma _{0}^{k} $ form an inverse system of Lie groups and
we can find a subgroup $K\subset lim_{\leftarrow}\Gamma _{0}^{k}$
whose projection on every $\Gamma _{0}^{k} $ is a maximal compact
subgroup for $k>0$. This follows from the fact that the kernel of the
projection $\Gamma _{0}^{k+1}\rightarrow \Gamma _{0}^{k}$ is a vector
space for $k>0$. The subgroup K is unique up to conjugation and its Lie
algebra k can be identified with a subalgebra of $a(\Gamma )$.

For our purposes we need the cohomology of basic elements rel K in
$a(\Gamma )$, namely $H(a(\Gamma );K)$ which is defined as follows:
Let $A\{a^k(\Gamma )\}$ denote the algebra of multilinear alternating
forms on $a^k(\Gamma )$ and let $A\{a(\Gamma )\}$ be the direct limit
of the $A\{a^k(\Gamma )\}$. The bracket in $a(\Gamma )$ induces a
differential on $A\{a(\Gamma )\}$ and we write $H\{a(\Gamma )\}$ for
the resulting cohomology group. The relative group $H^{*}(a(\Gamma
);K)$ is now defined as the cohomology of the subcomplex of
$A\{a(\Gamma )\}$ consisting of elements which are invariant under the
natural action of K and annihilated by all inner products with
elements of k. Then the result is:

\textsl{Let $F$ be a $\Gamma $-foliation on M. There is an
algebra homomorphism}

$$\phi :H\{a(\Gamma );K\}\rightarrow H(M;{\bf R})$$

which is a natural transformation on the category $C(\Gamma )$.

The construction of $\phi $ is as follows:

Let $P^k(\Gamma )$ be the differential bundle of k-jets at the origin
of elements of $\Gamma $. It is a principal $\Gamma
_{0}^{k}$-bundle. On the other hand $\Gamma $ acts transitively on the
left on $P^k(\Gamma )$. Denote by $A(P^{\infty}(\Gamma ))$ the direct
limit of the algebras $A(P^{k}(\Gamma ))$ of differential forms on
$P^k(\Gamma )$. The invariant forms wrt the action of $\Gamma $
constitute a differential subalgebra denoted $A_{\Gamma }$. One can
then prove that it is actually isomorphic to $A(a(\Gamma ))$.

Now let F be a foliation on M and let $P^k(F)$ be the differentiable
bundle over M whose fibre at every point say $x\in M$ is the space of k-jets at
this point of local projections that vanish on $x$. This is a $\Gamma
_0^k $-principal bundle. Its restriction is isomorphic to the inverse
image of the bundle $P^k(\Gamma )$, hence the differential algebra of
$\Gamma $-invariant forms on $P^k(\Gamma )$ is mapped in the algebra
$A(P^k(F))$ of differential forms on $P^k(F)$. If we denote by
$A(P^{\infty}(F))$ the direct limit of $A(P^{k}(F))$ we get an
injective homomorphism $\phi $ of $A(a(\Gamma ))$ in $A(P^{\infty}(F
))$ commuting with the differential.

This homomorphism is compatible with the action of K, hence induces a
homomorphism on the subalgebra of K-basic elements. But the algebra
$A(P^{k}(F );K)$ of K-basic elements in $A(P^{k}(F))$ is isomorphic to
the algebra of differential forms on $P^k(F)/K$ which is a bundle over
M with contractible fibre $\Gamma _0^k/K$.Hence $H(A(P^k(F);K))$ is
isomorphic via the de Rham theorem to $H(M;{\bf R})$. The homomorphism
$\phi $ is therefore obtained as the composition

$$H(a(\Gamma );K)\rightarrow H(A(P^{\infty }(F);K))=H(M;{\bf R})$$

But we think that is enough with \emph{abstract nonsense}
formalism. Let us make our discussion more \textsl{down to earth}:

Consider the GDA (over {\bf R})

$$WO_{q}=\wedge (u_1, u_3,..., u_{2(q/2)-1})\otimes P_q(c_1,...,
c_q)$$

with $du_i=c_i $ for odd i and $dc_{i}=0$ for all i and
$$W_{q}=\wedge (u_{1},u_{2},...,u_{q})\otimes P_{q}(c_1,...,c_q)$$

with $du_{i}=c_{i}$ and $dc_{i}=0$ for i=1,...,q where $deg
u_{i}=2i-1$, $deg c_{i}=2i$ and $\wedge $ denotes exterior algebra,
$P_q $ denotes the polynomial algebra in the $c_{i}$'s mod elements of
total degree greater than 2q. The cohomology of $W_q $ is the Gelfand
Fuchs cohomology of the Lie algebra of formal vector fields in q
variables. We note that the ring structure at the cohomology level is
trivial, that is all cup products are zero. Then the main result is
that there are homomorphisms
$$\phi :H^{*}(WO_{q})\rightarrow H^{*}(B\Gamma ^{r}_{q};{\bf R})$$

$$\tilde {\phi }:H^{*}(W_q)\rightarrow H^{*}(\tilde{B\Gamma _{q}^{r}};{\bf
R})$$

for $r \geq 2$ with the following property ($\tilde{B\Gamma _{q}^{r}}$
denotes the classifying space for \emph{framed} foliations): If $F$ is a codim
q $C^{r}$ foliation of a manifold M, there is a GDA homomorphism
$$\phi _{F}:WO_{q}\rightarrow \wedge ^{*}(M)$$
into the de Rham algebra on M, defined in terms of the differential
geometry of $F$ and unique up to chain homotopy, such that on
cohomology we have $\phi _{F}=f^{*}\circ \phi $, where
$f:M \rightarrow B\Gamma ^{r}_{q}$ classifies $F$. If the
normal bundle of $F$ is trivial, there is a homomorphism
$$\tilde {\phi _{F}}:W_q \rightarrow \wedge ^{*}(M)$$
with analogous properties.\\

Combining this result with the fact that $B\tilde{\Gamma ^{0}_{q}}$ is
\emph{contractible}, we deduce that a foliation is essentially
determined by the structure of its normal bundle; the \emph{Chern}
classes of the normal bundle are contained in the image of the map
$\phi $ above but we have \emph{additional} non trivial classes in the
case of foliations (which are rather difficult to find though), one of
which is this BHGV class which we
constructed explicitly and it is the class we use as a Lagrangian
density which is purely topological since its degree fits nicely for
describing 5-branes.

There is an alternative approach due to Simons \cite{simons} which avoids passing to
the normal bundle using circle coefficients. What he actually does is
to associate to a principal bundle with connection a family of
characteristic homomorphisms from the integral cycles on a manifold to
$S^{1}$ and then defining an extension denoted $K^{2k}_{q}$ of
$H^{2k}(BGL_{q};{\bf Z})$. This approach is related
to \emph{gerbes}. A gerbe over a manifold is a construction which
locally looks like the Cartesian product of the manifold with a line
bundle. Clearly it is a special case of foliations (remember our
previous comment on foliations). However this
approach actually suggests that they might be equivalent, if the
approach of Bott-Haefliger is equivalent to that of Simons,
something which is not known.\\

\textsl{Now the conjecture is that the \emph{partition function} of this
Lagrangian is related to the invariant introduced in \cite{Z2}.}\\

In order to establish relation with physics, we must make some
identifications. The 1-form $\theta $ appearing in the Lagrangian has
no direct physical meaning. In physics it is assumed that a 5-brane
gives rise to a 6-form gauge field denoted $A_6$ whose field strength
is simply
\begin{equation}
dA_{6}=F_{7}
\label{eq3}
\end{equation}
The only way we can explain geometrically this is that this 6-form is
the Poincare dual of the 6-chain that the 5-brane sweps out as it
moves in time. 

We know that since we have S-duality between membranes and 5-branes,
in an obvious notation one has
\begin{equation}
F_{7}=*F_{4}
\label{eq4}
\end{equation}
which is the S-duality relation, where
\begin{equation}
F_{4}=dA_{3}
\label{eq5}
\end{equation}

Observe now that the starting point for 5-brane theory is $A_6$ where
the starting point to construct the BHGV class was the 5-form $\Omega
$. How are they related?

There are three obvious possibilities:

I. $d\Omega =A_6$
That would imply that $A_6 $ is pure gauge.

II.$dF_{4}=\Omega $
This is trivial because it implies $d\Omega =0$, hence $d\Omega
=\theta \wedge \Omega =0$.

III. The only remaining possibility is
\begin{equation}
*A_{6}=\Omega
\label{eq6}
\end{equation}

We call this \emph{``reality condition''}. So now in principle we can substitute
equations (10) and (5) into (6) and get an expression for the Lagrangian
which involves the gauge field $A_6$.

The Euler-Lagrange equations which are actually analogous to D=11 N=1
supergravity Euler-Lagrange equations (see equation (12) below) read:
\begin{equation}
d*d\theta + \frac{1}{5}(d\theta )^{5}=0
\label{eq7}
\end{equation}

The on-shell relation with D=11 N=1 supergravity is established as
follows: recall that the bosonic sector of this supergravity theory is
$$\int F_{4}\wedge F_{4}\wedge A_{3}$$
 where $F_{4}=dA_{3}$ with
Euler-Lagrange equations
\begin{equation}
d*F_{4}+\frac{1}{2}F_{4}\wedge F_{4}=0
\label{eq8}
\end{equation}

Constraining $A_3 $ via (12), by (9), (8), (7), (10) and (5) we get a constraint
for $\theta $ which can be added to the class $\gamma $ as a Lagrange
multiplier.

In order to calculate the partition function, some additional
difficulties may arise because we do not know what notion of
equivallence between foliations is the appropriate one for physics in
order to fix the gauge and add Faddeev-Popov terms as constraints to
kill-off the gauge freedom. There are actually four different notions
of equivalence for foliations: conjugation, homotopy, integrable
homotopy and foliated cobordism.

In principle, one must end up with an equivallent theory starting with
membranes (that's due to S-duality), provided of course a suitable
class was found. Clearly the BHGV class for a membrane would be a
\emph{17-form}. 

The final comment refers to \cite{sus}. In that article it was conjectured
that the quantum mechanics of branes could be described as a matrix model.
As it is well-known matrix models use point particle degrees of freedom.
This is rather intriguing since we are talking about M-theory which
contains various p-branes. In our approach though we propose a Lagrangian
density which has as fundamental object a mysterious 1-form which, if seen
as a gauge potential, that  would imply the existence of some yet unknown
underlying point particle!

\subsection{Plane fields}

We now pass on to the second question raised in this application, namely the
restrictions on the topology of the underlying manifold of a theory
containing p-branes via our physical principle.

It is clear from the definition that the existence of a 
foliation  of certain dim, say d (or equivalently codim q=n-d) on an
n-manifold (closed) depends:

a.) On the existence of a dim d subbundle of the tangent bundle

b.) On this d-dim subbundle being integrable.

The second question has been answered almost completely by Bott and in
a more general framework by
Thurston. Bott's result dictates that for a codim q subbundle of the
tangent bundle to be integrable, the ring of Pontrjagin classes of the
subbundle with degree $>2q$ must be zero. There is a secondary
obstruction due to Shulman involving certain Massey triple products
but we shall not elaborate on this. However Bott's result suggests
nothing for question a.) above. Let us also mention that this result of
Bott can be deduced by another theorem due to Thurston which states
that the classifying space $B\tilde{\Gamma ^{\infty }_{q}}$ of smooth
codim q framed foliations is (q+1)-connected. 

On the contrary, Thurston's result reduces the existence
of codim $q>1$ foliations (at least up to homotopy) to the existence of
\emph{q-plane fields}. This is a deep question in differential
topology, related to the problem of classification of closed manifolds
according to their \textsl{rank}.

Now the problem of existence of q-plane fields has been answered
\emph{only for some cases for spheres $S^n$ for various values of
n,q} \cite{steenrod}.
In particular we know everything for spheres of dimension 10 and
less. We should however mention a theorem due to Winkelnkemper
\cite{win1} which
is quite general in nature and talks about simply connected compact
manifolds of dim n greater than 5. If n is not 0 mod 4 then it admits
a so-called \emph{Alexander decomposition} which under special
assumptions can give a particular
kind of a codim 1 foliation with $S^{1}$ as space of leaves and a
surjection from the manifold to $S{1}$. If n is 0 mod 4 then the manifold admits an
Alexander decomposition iff its signature is zero.

Let us return to string theory now: String theory works in D=10 and in this case we have the old brane-scan
suggesting the string/5-brane duality. The new brane-scan contains all
p-branes for $p\leq 6$ and some D-7 and 8-branes are thought to
exist. However topology says that for a sphere in dim 10 we can have
only dim 0 and dim 10 plane fields (in fact this is true for all even
dim spheres), hence by Thurston only dim 0 and
dim 10 foliations and then our physical principle suggests that $S^{10}$
is ruled out as a possible underlying topological space for string theory.\\

What about M-Theory in D=11 then?\\

For the case of $S^{11}$ then it is known that $S^{11}$ admits a 3-plane field, hence by our physical
principle a theory containing membranes \emph{can} be formulated on
$S^{11}$. For $S^{11}$ nothing is known for the existence of q-plane
fields for q greater than 3. But now we apply S-duality between membranes/5-branes
 and conjecture that:\\

\emph{$S^{11}$ should admit 5-plane fields}.\\

Let us close with two final remarks: 

1. There is extensive work in
foliations with numerous results which actually insert many extra
parameters into their study, for example metric aspects, existence of
foliations with compact leaves (all or at least one or exactly one),
with leaves diffeomorphic to ${\bf R^n}$ for some n etc. We do not have
a clear picture for the moment concerning imposing these in physics.
Let us only mention
one particularly strong result due to Wall generalising a result of
Reeb \cite{reeb}: if a closed
n-manifold admits a codim 1 foliation whose leaves are homeomorphic to
${\bf R^{n-1}}$, then by Thurston we know that its Euler characteristic
must vanish, but in fact we have more: it has to be the n-torus!\\

The interesting point
however is that although all these extended objects theories in
physics are expressed as $\sigma $ models \cite{Z1}, hence they
involve metrics on the manifold (target space) and on the worldvolumes ie on the
leaves, in our approach the metric is only used in the reality
condition (10) which makes connection with physical fields (that is
some metric on the target space) where at the same time we do not use any
metric on the source space (worldvolumes-leaves of the foliation).\\

2. In \cite{Z1} another Lagrangian density was proposed. It is
different from the one described here but they are related in an
analogous way to the relation between the Polyakov and Nambu-Goto (in
fact Dirac \cite{dirac}) actions for the free bosonic string: extended
objects basically immitate string theory and we have two formalisms:
the $\sigma $ model one which is the Lagrangian exhibited in \cite{Z1}
using Polyakov's picture of $\sigma $ models as flat principal bundles
with structure group the isometries of the metric on the target space
\cite{polyakov}; yet we also have the \emph{embedded surface} picture
which is the Dirac (Nambu-Goto) action and whose analogue is described
in this work.

In the light of a very recent work \cite{WIT}, we can also make some
further comments: the first is that the Moyal algebra used in order to
discuss noncommutative solitons is actually Morita equivalent to the usual
commutative one. This fact can be further verified from the explicit
construction of an algebra homomorphism between noncommutative and
ordinary Yang-Mills fields based on Gelfand's theorem. Truly
noncommutative situations appear when discussing noncommutative tori.

The next comment refers to the last section of that paper: we already know
that
strings in a constant magnetic field can be described from that Moyal
algebra-like spacetime structure and they also discuss what may happen to
5-branes in M-theory. Our point of view coincides with theirs in the
following way: we here propose that this is indeed the case for
5-branes with a C-field turned on, namely that this situation can be
described by 6-dim foliations which rather correspond to a "free" theory
of
branes but on a "noncommutative" topological space, which is actually, in
our case, the space of leaves of the corresponding foliation.

\end{document}